\documentclass[12pt]{iopart}

\usepackage{xcolor} 
\usepackage{graphicx}
\usepackage{dcolumn}
\usepackage{bm}
\usepackage{tabularx}
\usepackage{rotating}
\usepackage{makecell}
\usepackage{float}
\usepackage{multirow}

\begin{document}

\title{Investigation of the magnetic ground state of GaV$_4$S$_8$ using powder neutron diffraction}

\author{S. J. R. Holt$^1$, C. Ritter$^2$, M. R. Lees$^1$, and G. Balakrishnan$^1$}
\address{$^1$ University of Warwick, Department of Physics, Coventry, CV4 7AL, United Kingdom}
\address{$^2$ Institut Laue-Langevin, 71 Avenue des Martyrs, CS20156, 38042 Grenoble C\'edex 9, France}
\ead{S.J.R.Holt@warwick.ac.uk, G.Balakrishnan@warwick.ac.uk}

\begin{abstract}
The magnetic ground state of polycrystalline N\'eel skyrmion hosting material GaV$_4$S$_8$ has been investigated using \textit{ac} susceptibility and powder neutron diffraction.
In the absence of an applied magnetic field GaV$_4$S$_8$ undergoes a transition from a paramagnetic to a cycloidal state below 13~K and then to a ferromagnetic-like state below 6~K.
With evidence from \textit{ac} susceptibility and powder neutron diffraction, we have identified the commensurate magnetic structure at 1.5~K, with ordered magnetic moments of $0.23(2)~\mu_{\mathrm{B}}$ on the V1 sites and $0.22(1)~\mu_{\mathrm{B}}$ on the V2 sites.
These moments have ferromagnetic-like alignment but with a 39(8)$^{\circ}$ canting of the magnetic moments on the V2 sites away from the V$_4$ cluster.
In the incommensurate magnetic phase that exists between 6 and 13~K, we provide a thorough and careful analysis of the cycloidal magnetic structure exhibited by this material using powder neutron diffraction.
\end{abstract}

\noindent{\it Keywords\/}: Skyrmion Materials, Powder Neutron Diffraction, Magnetism, Magnetic Structure, Cycloidal Magnetic Structure

\submitto{\JPCM}

\maketitle
\ioptwocol

\section{Introduction}
There has been widespread interest in skyrmion hosting materials, both for their fascinating magnetic properties, as well as their potential application in fields such as spintronics \cite{fert2013skyrmions}.
These topological magnetic quasi-particles have been found in an increasing number of systems but the majority of these materials stabilise Bloch as opposed to N\'eel skyrmions.
In order to explain the physics of N\'eel skyrmions we need a deeper understanding of the materials in which they form, including the various types of magnetism that these systems exhibit \cite{lancaster2019skyrmions}.

\begin{figure}[t]
\centering
\includegraphics[width=0.8\linewidth]{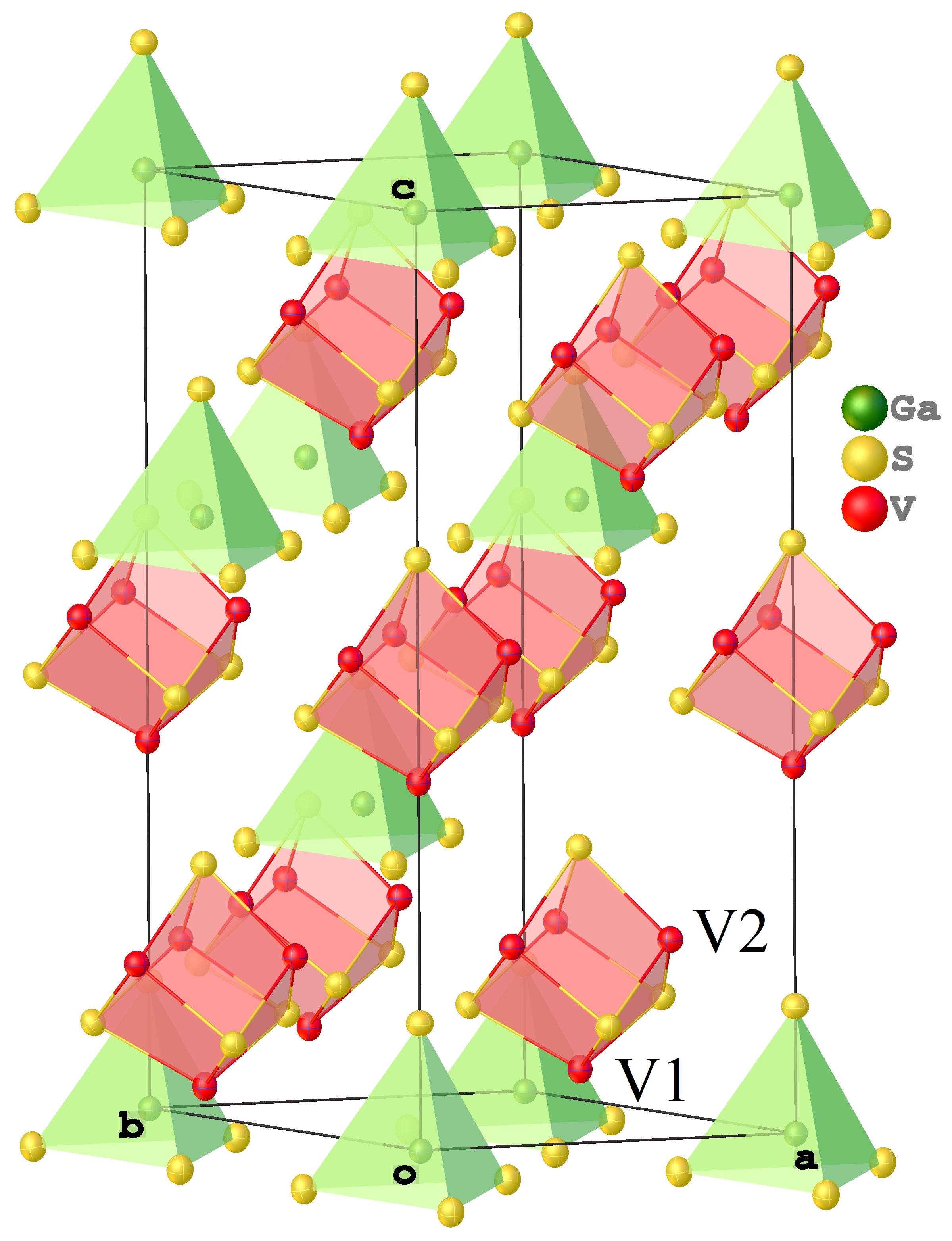}
\caption{Low temperature $R3m$ crystal structure in the hexagonal setting with the V$_4$S$_4$ heterocubane shown in red and the GaS$_4$ tetrahedra shown in green.
The V atoms are shown in red, Ga in green, and S in yellow.}
\label{Fig:crystal_struc}
\end{figure}

GaV$_4$S$_8$ is a member of the lacunar spinel family ($AB_4X_8$) which has been shown to stabilise a variety of magnetic phases including a cycloidal state  and N\'eel skyrmions \cite{kezsmarki2015neel}. 
It adopts a high-temperature face centred cubic $F\overline{4}3m$ structure with a lattice made up of heterocubane-like $[\textrm{V}_4\textrm{S}_4]^{5+}$ and $[\textrm{GaS}_4]^{5-}$ tetrahedra \cite{stefancic2020establishing}.
A first-order structural phase transition occurs at 42~K distorting the cubic lattice into a low-temperature rhombohedral $(R3m)$ crystal structure.
In this paper we use the hexagonal setting of the $R3m$ space group.
This phase transition causes a deformation of the V$_4$ tetrahedra, elongating along one of the four possible $\langle 111 \rangle$ cubic directions, creating two distinct V sites.
Figure~\ref{Fig:crystal_struc} depicts the low temperature crystal structure of GaV$_4$S$_8$.
The V1 (3\textit{a} Wyckoff) site corresponds to the V ion on the elongated node of the tetrahedra while the V2 (9\textit{b} Wyckoff) site corresponds to the three remaining symmetric V ions within the tetrahedra. 
As this distortion can occur in any of the four possible $\langle 111 \rangle$ directions, micro-domains of each orientation are formed.

\begin{figure}[t]
\centering
\includegraphics[width=1\linewidth]{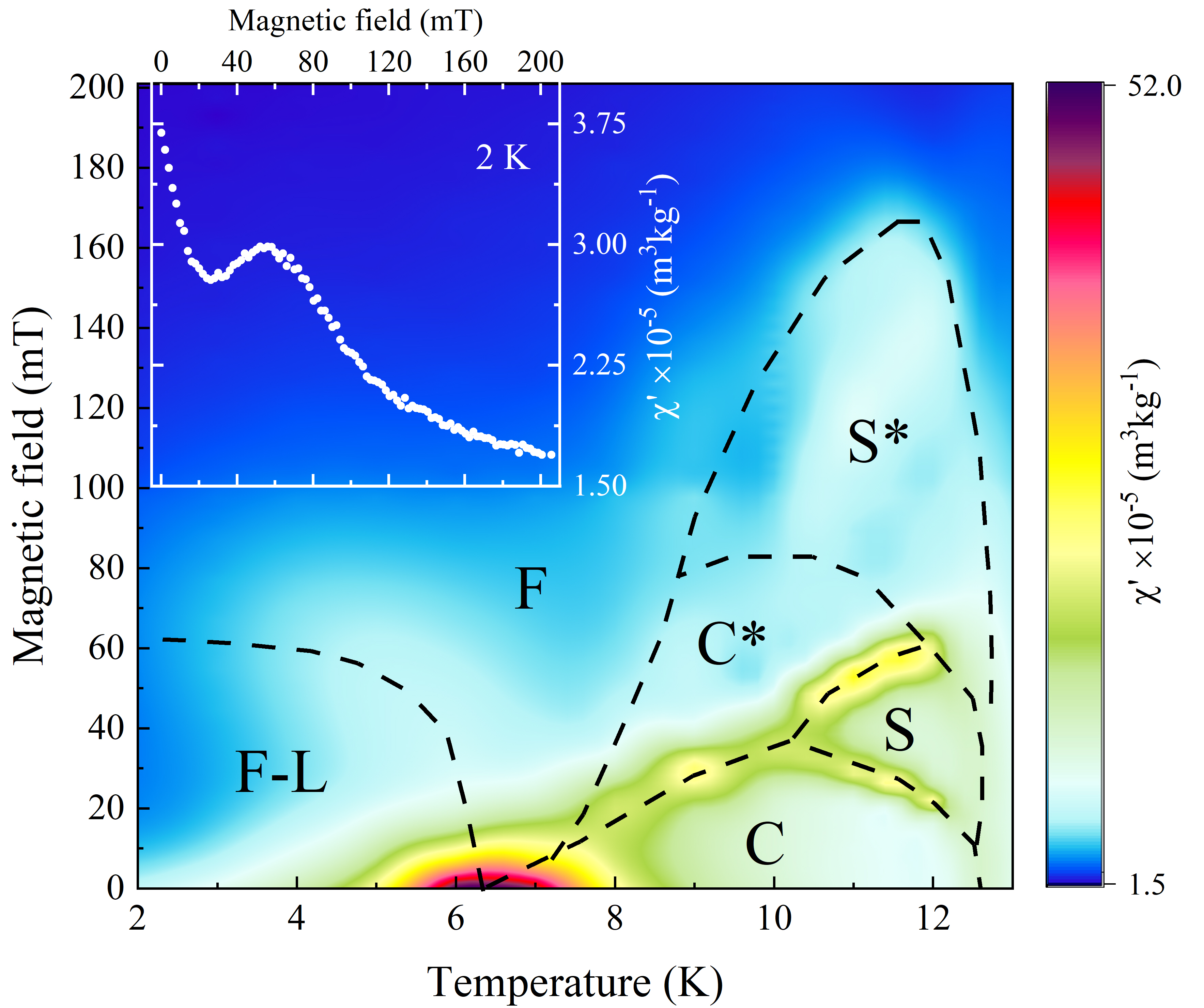}
\caption{Real component of the \textit{ac} susceptibility $(\chi')$ versus temperature and applied \textit{dc} field for a single crystal of GaV$_4$S$_8$ with the \textit{ac} and \textit{dc} magnetic fields applied along the pseudocubic [111] direction showing the low and high field skyrmion phases (S and S*), the low and high field cycloidal phases (C and C*), ferromagnetic-like phase (F-L), and ferromagnetic phase (F).
Dashed lines are given as guides to the eye.
The inset depicts an \textit{ac} susceptibility field scan at 2~K.}
\label{Fig:AC}
\end{figure}

\begin{table}[t]
\centering
\caption{Basis vectors (BV) for GaV$_4$S$_8$ used in the magnetic refinements for each propagation vector.}
\label{tab:BV}
    \begin{tabularx}{0.9\linewidth}{>{\centering\arraybackslash}X >{\centering\arraybackslash}X >{\centering\arraybackslash}X >{\centering\arraybackslash}X}
        \hline
        \hline
        & BV1 & BV2 & BV3\\
        \hline
        \multicolumn{4}{c}{$\bm{k}=\bm{0}$}\\
        \\
        \multicolumn{4}{c}{V1 (3a)} \\
        x,y,z & 0 0 1 \\
        \multicolumn{4}{c}{V2 (9b)} \\
        x,y,z & 1 2 0& 0 0 1 \\
        -y,x-y,z & -2 -1 0 & 0 0 1 \\
        -x+y,-x,z & -1 -1 0 & 0 0 1 \\
        \hline
        \multicolumn{4}{c}{$\bm{k}_{0k0}=[0K0]$}\\
        \\
        \multicolumn{4}{c}{V1 (3a)} \\
        x,y,z & 1 2 0& 0 0 1 \\
        \multicolumn{4}{c}{V2$_1$ (9b)} \\
        x,y,z & 1 2 0 & & 0 0 1 \\
        \multicolumn{4}{c}{V2$_2$ (9b)} \\
        x,y,z & 1 0 0 & 0 1 0 & 0 0 1 \\
        \mbox{-x+y,y,z} & -1 0 0& 1 1 0 & 0 0 1 \\
        \hline
        \multicolumn{4}{c}{$\bm{k}_{00l}=[00L]$}\\
        \\
        \multicolumn{4}{c}{V1 (3a)} \\
        x,y,z & 0 0 1 \\
        \multicolumn{4}{c}{V2$_1$, V2$_2$, and V2$_3$(9b)} \\
        x,y,z & 1 2 0 & 0 0 1 \\
        -y,x-y,z & -2 -1 0 & 0 0 1 \\
        \mbox{-x+y,-x,z} & 1 -1 0 & 0 0 1 \\
        \hline
        \multicolumn{4}{c}{$\bm{k}_\textrm{ip}$ and $\bm{k}_\textrm{ip-c}$}\\
        \\
        \multicolumn{4}{c}{V1 (3a), V2$_1$, V2$_2$, and V2$_3$(9b)}\\
        x,y,z & 1 0 0 & 0 1 0 & 0 0 1 \\
        \hline
        \hline
    \end{tabularx}
\end{table}

\begin{table*}[t]
\centering
\caption{Parameters refined for the magnetic models of GaV$_4$S$_8$ at 1.5 and 9~K using the FullProf software suite \cite{Rodriguez1993recent}. $\lambda$ is the magnetic modulation length calculated from the refined propagation vector ($\bm{k}$ vector) using $\lambda^2 = (a/H)^2 + (b/K)^2 + (c/L)^2$. For simplicity, the BVs of $\bm{k}_\textrm{ip}$ and $\bm{k}_\textrm{ip-c}$ are given for propagation along the $0K0$ direction but this can be generalised to any in-plane direction by taking the appropriate ratios of BV1 and BV2.}
\label{tab:Mag}
\resizebox{\textwidth}{!}{
    \begin{tabular}{cccc|cccc|cccc|c}
        \hline
        \hline
        &&&& \multicolumn{4}{c|}{V1} & \multicolumn{4}{c|}{V2}&\\
        Temperature & $\bm{k}$ vector & $\chi ^2$ & $R_\textrm{mag} $ & BV1 & BV2 & BV3 & Total moment $\left(\mu_{\mathrm{B}}\right)$& BV1 & BV2 & BV3 & Total moment $\left( \mu_{\mathrm{B}}\right)$ & $\lambda$ (nm)\\
        \hline
        \multirow{2}{*}{1.5 K} & $\bm{0}$ & 5.11 & 15.31 & 0.23(2) & - & - & 0.23(2) & -0.079(8) & 0.17(1) & - & 0.22(1) & - \\
        & $\bm{0}$ & 5.11 & 15.13 & 0.38(1) & - & - & 0.38(1) & - & 0.12(1) & - & 0.12(1) & - \\
        \hline
        \multirow{4}{*}{9 K} & $[0K0]$ & 4.08 & 8.0 & - & 0.47(2) & - & 0.47(2) & - & - & 0.15(2) & 0.15(2) & 38(2)\\
        & $[00L]$ & 4.59 & 15.65 & 1.00(2) & - & - & 1.00(2) & - & 0.10(1) & - & 0.10(1) & 21(1)\\
        & $\bm{k}_\textrm{ip}$ & 4.30 & 11.4 & - & 0.341(9)\textsuperscript{*} & 0.34(1) & 0.34(1) & - & 0.12(1)\textsuperscript{*} & 0.12(1) & 0.12(1) & 36(2) \\
        & $\bm{k}_\textrm{ip-c}$ & 4.55 & 12.5 & - & 0.21(1)\textsuperscript{*} & 0.21(1)& 0.21(1) & - & 0.21(1)\textsuperscript{*} & 0.21(1) & 0.21(1) & 32(2)\\
        \hline
        \hline
    \end{tabular}
    }
    \textsuperscript{*} Imaginary BV creating a 90 degree phase difference between the two BVs.
\end{table*}

The magnetic behaviour of GaV$_4$S$_8$ arises from the sharing of a single unpaired electron among the V ions within the V$_4$ tetrahedron similar to other members of the lacunar spinel family \cite{pocha2000electronic}. 
GaV$_4$S$_8$ undergoes a paramagnetic to cycloidal magnetic transition at 13~K in zero field.
Application of a magnetic field to the cycloidal magnetic phase stabilises magnetic skyrmions \cite{kezsmarki2015neel}. 
A further reduction in temperature produces a ferromagnetic-like state below 6~K \cite{stefancic2020establishing, holt2020structure}.
Due to the easy axis anisotropy, the direction of the applied field with respect to the crystal lattice strongly effects the magnetic phases present.
For example, aligning the field along the [111] pseudo-cubic crystallographic direction has been found to stabilise the largest extent of the skyrmion phase in magnetic field-temperature space \cite{bordacs2017equilibrium}.

The magnetic properties of GaV$_4$S$_8$ have previously been investigated using a variety of techniques such as muon-spin rotation ($\mu^+$SR) \cite{franke2018magnetic, hicken2020magnetism}, small angle neutron scattering (SANS) \cite{kezsmarki2015neel, white2018direct, dally2020magnetic}, and electric polarisation \cite{widmann2017multiferroic}.
Magnetisation measurements \cite{widmann2017multiferroic} have shown that at temperatures below 6 K and low fields, the magnetic structure of GaV$_4$S$_8$ is not a simple collinear FM phase as suggested by previous work \cite{kezsmarki2015neel}.
High temperature magnetisation measurements have also emphasised the magneto-structural link in this material, with discontinuities being seen in the inverse susceptibility at the structural phase transition \cite{widmann2017multiferroic, stefancic2020establishing}.
Both the zero-field magnetic and structural transitions are able to be seen in heat capacity measurements, while the structural transition is also evident in resistivity data \cite{widmann2017multiferroic}.

Our previous powder neutron diffraction studies on the ferromagnetic-like state revealed a magnetic spin texture where the moments on the V1 sites align with the hexagonal crystallographic $c$ axis and the moments on the V2 sites cant outwards from the centre of the V$_4$ cluster \cite{stefancic2020establishing}.
Investigations of GaV$_4$S$_8$ support the evidence for a commensurate magnetic ground state below 6~K and a cycloidal magnetic phase has been identified to propagate along the $[1\overline{1}0]$ pseudo-cubic direction from SANS measurements, although due to complex micro-twinning, the propagation vector of the magnetism with respect to the rhombohedral phase has not been reported to date \cite{kezsmarki2015neel, white2018direct}.

We have therefore carried out \textit{ac} susceptibility measurements on a single crystal of GaV$_4$S$_8$ and have identified a new commensurate ferromagnetic-like zero-field phase at low temperatures that undergoes a transition to a ferromagnetic state with applied magnetic fields.
In order to investigate this ferromagnetic-like region in more detail, we have performed high-intensity powder neutron diffraction experiments on GaV$_4$S$_8$.
We describe the magnetic structure in this low temperature commensurate phase and the incommensurate state, with reference to the existing literature.

\section{Experimental Details}
Polycrystalline samples of GaV$_4$S$_8$ were synthesised by solid-state reaction as described in Ref.~\cite{stefancic2020establishing}.
Phase purity was confirmed by laboratory powder X-ray diffraction using monochromatic Cu K$_{\alpha 1}$ radiation.
Powder neutron diffraction experiments were carried out on the D2B and D20 diffractometers at the Institute Laue-Langevin (ILL), in line with our previous experiments on these diffractometers \cite{stefancic2020establishing, holt2020structure}.
A standard orange cryostat was used for neutron diffraction with the $\sim 9.5$~g powder sample sealed in a thin-walled cylindrical vanadium tube of 12~mm diameter.
Magnetic structure determination measurements were performed on the D20 diffractometer with a neutron wavelength of $2.417$~\AA\ at three temperatures: 1.5, 9, and 15~K representing the lowest temperature achievable in the cryostat in the ferromagnetic-like region, in the centre of the cycloidal phase, and in the paramagnetic phase, respectively. 
The magnetic refinements were carried out using the FullProf software suite \cite{Rodriguez1993recent}.
\textit{ac} susceptibility measurements as a function of \textit{dc} applied field were performed on a single crystal of GaV$_4$S$_8$ grown as described in Ref.~\cite{stefancic2020establishing}.
An \textit{ac} field of 113~Hz and 0.3~mT was applied after zero-field-cooling then application of a \textit{dc} magnetic field along the cubic [111] direction at intervals of approximately of 0.25~K and 2~mT.

\section{Results and Discussion}
\subsection{\label{sec:ac} \textit{ac} susceptibility}
Figure \ref{Fig:AC} depicts the \textit{ac} susceptibility phase diagram obtained on a GaV$_4$S$_8$ single crystal.
Comparison with the magnetic phase diagram by K\'ezsm\'arki \textit{et al}. \cite{kezsmarki2015neel} enables identification of the low field cycloidal (C) and skyrmion (S) phases and the high field cycloidal (C*) and skyrmion (S*) phases.
However, an additional low temperature ferromagnetic-like (F-L) phase can be identified from the \textit{ac} susceptibility map in Fig.~\ref{Fig:AC}, which is not present in most previously published phase diagrams. 
Based on SANS measurements from the literature \cite{kezsmarki2015neel, white2018direct, dally2020magnetic} and our previous powder neutron diffraction measurements \cite{stefancic2020establishing}, this low temperature magnetic state has been confirmed to be a commensurate structure.
This is commonly identified in the literature as a ferromagnetic state although more careful \textit{ac} susceptibility measurements in the low-temperature low-field region shows a peak in the susceptibility with application of a \textit{dc} magnetic field that extends from 6~K down to the lowest temperatures measured, as seen in Fig.~\ref{Fig:AC}. 
The F-L phase had not been identified in previous studies using techniques such as SANS and atomic force microscopy \cite{kezsmarki2015neel} that are often not sensitive to small changes in a commensurate magnetic structure, although a non-collinear magnetic state, possibly  with short-range order, was suggested from magnetisation measurements \cite{widmann2017multiferroic}.
The inset of Fig.~\ref{Fig:AC} highlights this peak at 2~K in the \textit{ac} susceptibility which is present at $\sim 65$~mT on decreasing field after saturation at 1~T.
As this peak is observed after both zero-field-cooling and saturation with a high magnetic field, it precludes the peak's origin being from magnetic domains and thus indicates a change in the dynamics of magnetic moments that is often associated with a magnetic phase transition.
The high-field low-temperature magnetic phase can be associated with a ferromagnetic state as it has previously been seen to be commensurate and no additional phase transitions have been observed with the application of larger magnetic fields.
Powder neutron powder diffraction in this ferromagnetic-like region, can help us further understand this magnetic phase and allow us to determine its magnetic structure

\begin{figure}[t]
\centering
\includegraphics[width=0.9\linewidth]{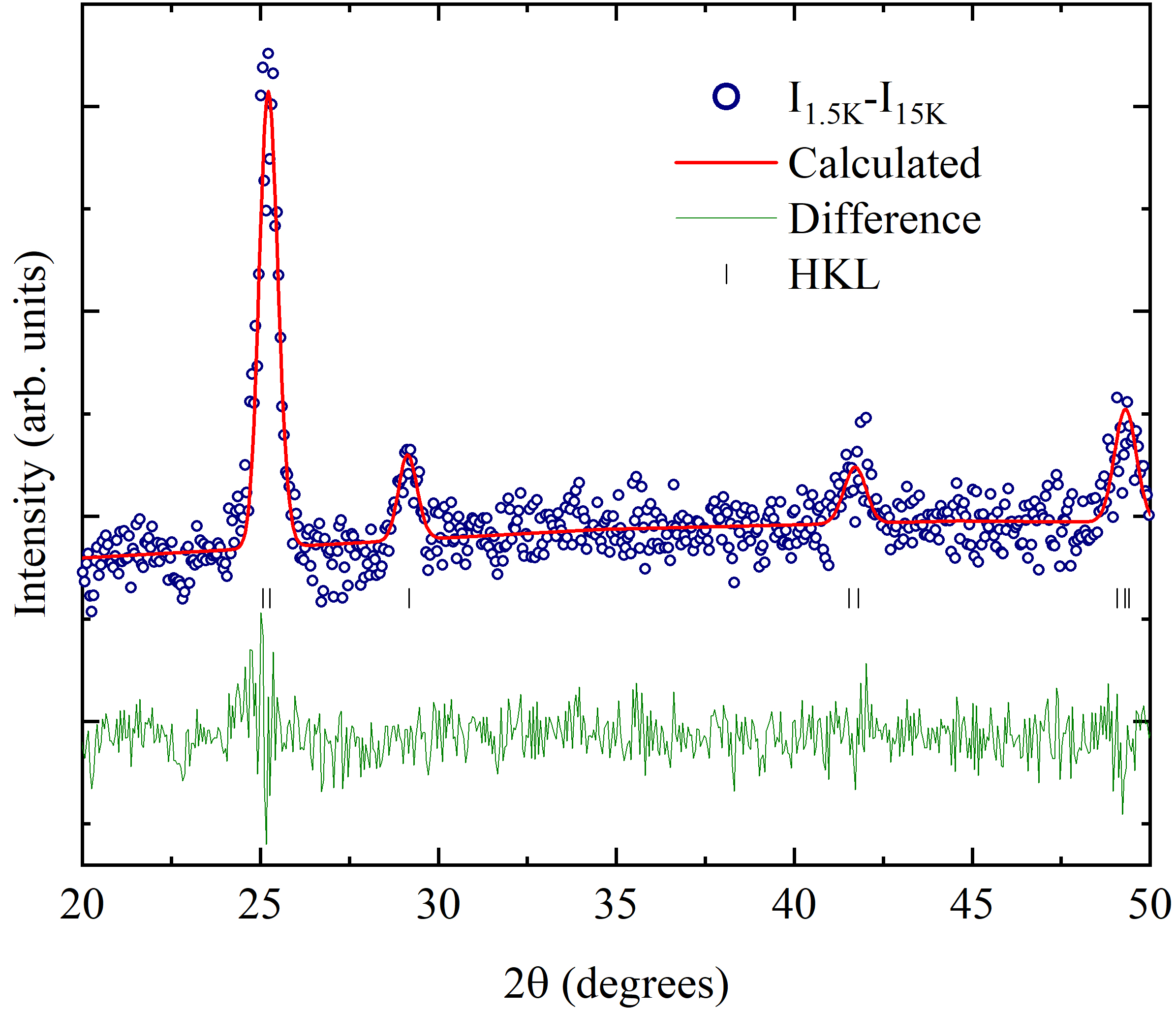}
\includegraphics[width=0.7\linewidth]{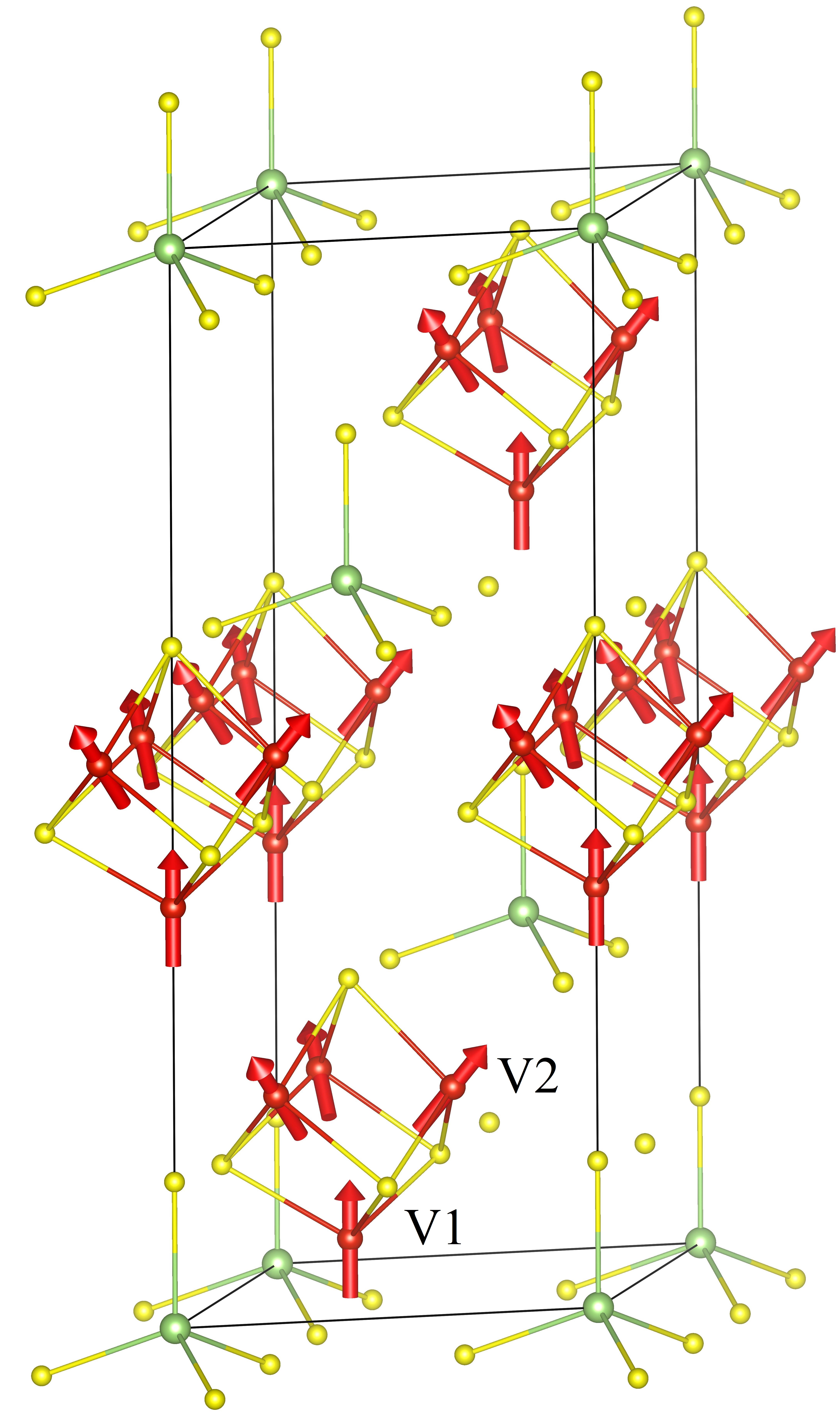}
\caption{Top: Experimentally-obtained magnetic difference pattern $1.5~\textrm{K}-15~\textrm{K}$ (blue open circles), predicted peak positions (black tick marks), magnetic Rietveld refinement (red solid line), difference (olive green solid line).
Data were taken on the D20 diffractometer.
Bottom: Magnetic structure of GaV$_4$S$_8$ at 1.5~K determined from powder neutron diffraction difference pattern refinements.}
\label{Fig:1p5K}
\end{figure}

\subsection{Powder Neutron Diffraction at 1.5~K: Commensurate Structure}
Figure~\ref{Fig:1p5K} shows the magnetic peaks in the $1.5~\textrm{K}-15~\textrm{K}$ powder neutron diffraction difference pattern.
The nuclear $R3m$ hexagonal structure was first refined at 15~K, a temperature at which GaV$_4$S$_8$ is paramagnetic.
At 1.5~K the peaks can be indexed by a $\bm{k}=\bm{0}$ propagation vector, in agreement with literature \cite{kezsmarki2015neel, stefancic2020establishing, dally2020magnetic}.
Magnetic symmetry analysis was performed using BASIREPS to determine the magnetic irreducible representations (IR) and their basis vectors \cite{rodriguez2010program, ritter2011neutrons}. 
The only IR that can be found to fit the data is $\Gamma_2$ for both the V1 and V2 sites. The corresponding basis vectors are given  in Table~\ref{tab:BV} and details of these refinements in Table~\ref{tab:Mag}.
The refinement yields a magnetic structure shown in Fig.~\ref{Fig:1p5K} with a total moment of $0.23(2)~\mu_{\mathrm{B}}$ on the V1 sites and $0.22(1)~\mu_{\mathrm{B}}$ on the V2 sites.
The V2 moments are canted outwards from the V$_4$ cluster, as seen in Fig.~\ref{Fig:1p5K} with total moment of the cluster being $0.88(5)~\mu_{\mathrm{B}}/$V$_4$. 
The magnetic structure is in agreement with our previous neutron diffraction results \cite{stefancic2020establishing} although the data presented in this paper show an equal moment distributed across all of the V atoms and have allowed us to determine a canting of the moments on the V2 sites of 39(8)$^{\circ}$, indicating that the low-temperature magnetic phase identified in the ac susceptibility measurements is not fully collinear.
In a recent SANS study, Dally \textit{et al.} suggested that the moments are constrained to point along the magnetic easy axis (the \textit{c} axis in the hexagonal setting of the $R3m$ space group) at 3~K \cite{dally2020magnetic}.
Constraining the refinement of our data in this way leads to nearly identical simulated pattern and $R$ factor, (see Table~\ref{tab:Mag}) but with a moment of $0.382(13)~\mu_{\mathrm{B}}$ on the V1 sites and $0.117(12)~\mu_{\mathrm{B}}$ on the V2 sites, hence to a total moment per cluster of $0.73(4)~\mu_{\mathrm{B}}/$V$_4$.
The unconstrained fit at 1.5~K gives better agreement with our previous magnetisation results \cite{stefancic2020establishing} along with the magnitude of ordered moment per cluster calculated by Dally \textit{et al.} \cite{dally2020magnetic}.
The slight difference of magnetic structure determined from our measurements and those of Dally \textit{et al.} can be attributed to how similar the diffraction patterns of the canted and collinear state are, making it difficult to distinguish between them.
It is worth noting that our observed pattern could not be fitted with a moment only residing on a single atom.

These zero-field measurements show a canting of the V2 spins away from the hexagonal \textit{c} axis, which is along the pseudocubic [111] direction.
This is consistent with the \textit{ac} susceptibility measurements that show the application of a \textit{dc} magnetic field along this [111] direction leads to a change in orientation of the magnetic moments from the canted to a collinear ferromagnetic structure as the field is increased.
The peak in $\chi$' suggests this reorientation occurs at a well-defined field (a phase boundary) rather than via a more gradual rotation of the magnetic moments.

\begin{figure}[t]
\centering
\includegraphics[width=0.9\linewidth]{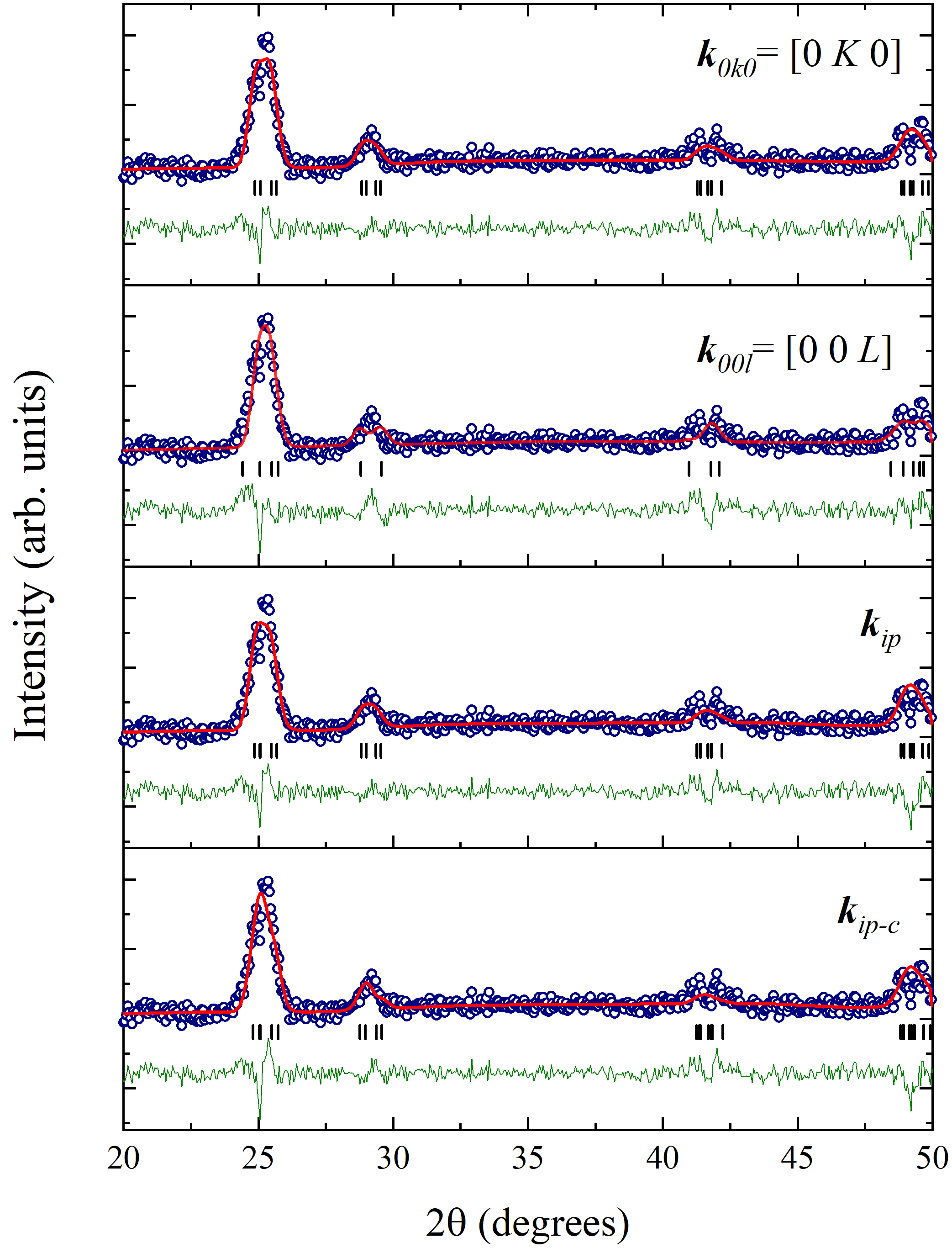}
\caption{Powder neutron diffraction of GaV$_4$S$_8$ fit using multiple incommensurate models with each propagation vector shown in a separate panel.
Experimentally-obtained magnetic difference pattern $9~\textrm{K}-15~\textrm{K}$ (blue open circles), predicted peak positions (black tick marks), magnetic Rietveld refinement (red solid line), difference (olive green solid line).
Data were taken on the D20 diffractometer.
}
\label{Fig:9K}
\end{figure}

\subsection{Powder Neutron Diffraction at 9~K: Incommensurate Structure}
Measurements carried out at 9~K within the incommensurate phase identified by SANS \cite{kezsmarki2015neel, white2018direct} (spanning from approximately 6 to 13~K) are shown in Fig.~\ref{Fig:9K}.
The magnetic Bragg peaks in the magnetic difference pattern can be seen to have broadened relative to 1.5~K, which highlights the incommensurate nature of the magnetism at 9~K.
Without previous knowledge of the form of magnetic structure, Occam's razor might be used to argue that the propagation vector would be along a primary crystallographic direction or the magnetic easy axis.
Refinements show that a solution using the $\bm{k}=0K0$ propagation vector with BV 001 (Table~\ref{tab:BV}), to be significantly better than any allowed $\bm{k}=00L$ refinement.
Due to symmetry the $\bm{k}=0K0$ propagation vector will give the same results as the $\bm{h}=H00$ propagation vector therefore only $\bm{k}=0K0$ is examined. 
The $\bm{k}=0K0$ refinement shown in Fig.~\ref{Fig:9K} also gives a visibly better fit than the $\bm{k}=00L$ solution.
However, the $\bm{k}=0K0$ refinement provides a solution that is a modulated sinusoidal wave propagating along the $\bm{k}=0K0$ direction with the magnetic moments aligned with the $c$ axis but a larger moment on the V1 site compared to the V2 sites.
The magnetic modulation length is slightly longer from the refinement compared to the value from SANS but due to the limitations of powder neutron diffraction for resolving magnetic structures with large length scales, this value seems reasonable \cite{kezsmarki2015neel}. 
The symmetry of this propagation vector precludes the possibility of cycloidal winding, going against existing evidence from measurements such as SANS \cite{kezsmarki2015neel, white2018direct}.

SANS results by White \textit{et al}. \cite{white2018direct} show the presence of a cycloidal winding (region C in Fig.~\ref{Fig:AC}) and they determine the magnetic propagation vector to be perpendicular to the hexagonal $c$ axis with no preferential direction in the plane.
As the propagation direction is ill defined but in the $ab$ plane, BASIREPS can be used to find a suitable IR and BV (Table~\ref{tab:BV}) for an arbitrary in-plane propagation vector $\bm{k}_\textrm{ip}$.
Using BV2 and BV3 with one real and one imaginary a cycloidal magnetic structure can be created.
From the literature \cite{white2018direct} we know there is a cycloidal magnetic winding in the plane of the propagation vector and the $c$ axis, therefore only solutions with this form of winding will be examined.
This yields two possible refinements, one with the moments on the distinct V Wyckoff sites being the same, and one where each type of site is refined separately and will be referred to as $\bm{k}_\textrm{ip-c}$ and $\bm{k}_\textrm{ip}$ respectively.
Both solutions give similar refinements, with the major difference being the size of the magnetic moment.
$\bm{k}_\textrm{ip-c}$ gives a good agreement with the commensurate state which indicated an approximate equal moment on each V site.

\section{Summary and Conclusions}
Using a detailed analysis of powder neutron diffraction data, combined with evidence from \textit{ac} susceptibility in the temperature region of interest, we have been able to establish that in low magnetic fields, a commensurate spin structure exists at 1.5~K with a canting of the magnetic spins as well as confirming cycloidal magnetic structure at 9~K.

At 1.5~K the magnetic structure determined is shown to be in agreement with our previous studies \cite{stefancic2020establishing}.
Taking into account the reported density functional theory calculations~\cite{hicken2020magnetism, dally2020magnetic}, results from our magnetisation measurements~\cite{stefancic2020establishing}, and the presence of a phase transition in the \textit{ac} susceptibility, the evidence supports a magnetic structure where the magnetic moments are, within error, spread equally across the V$_4$ cluster with a canting of the V2 spins.

In treating the incommensurate structure at 9~K, numerous models that differ only slightly in terms of the quality of the refinements, appear to be possible solutions.
The simplest solution suggests a sinusoidal magnetic structure with spin modulation along the $\bm{k}=0K0$ direction. 
However, this is inconsistent with previous SANS measurements of GaV$_4$S$_8$ which show a cycloidal magnetic structure propagating in the $ab$ plane whilst the moments can rotate between the $c$ axis and the propagation direction.
We have used this information to constrain the possible propagation directions in the $R3m$ phase for our refinements.
Fits then give a value for the modulation length close to that of SANS and two reasonable magnetic structures with the moments either the same on both the V1 and V2 sites, or varying by a factor of $\sim 3.5$.

This study highlights the need for further investigations, using techniques such as small angle neutron diffraction of materials that have the potential to host skyrmions which have been reported as having a sinusoidal modulating spin structures.
If, instead, the magnetic order in such materials show cycloidal winding this would leave open the possibility that these materials could be potential skyrmion hosts.

\section*{Acknowledgements}
This work is financially supported by the EPSRC UK Skyrmion Project Grant EP/N032128/1.
We are grateful for the beamtime allocated to us at the Institut Laue-Langevin (ILL) under the experiment code 5-31-2684.
Data are available from the ILL at DOI: 10.5291/ILL-DATA.5-31-2684.
We would like to thank O. A. Petrenko (University of Warwick) for their critical reading of the manuscript and T. J. Hicken (Durham University) for useful discussions.

\providecommand{\newblock}{}

\end{document}